\documentclass[final,5p,times,twocolumn]{elsarticle}

\usepackage[T1]{fontenc}
\usepackage[utf8]{inputenc}

\usepackage{amsmath}
\usepackage{amsfonts}
\usepackage{amssymb}
\usepackage{amsxtra}
\usepackage{array}
\usepackage{graphicx}
\usepackage{hepunits}
\usepackage{units}
\usepackage{color}
\usepackage{xspace}

\definecolor{purple}{rgb}{0.5,0,0.5}
\definecolor{blue}{rgb}{0.0,0,0.9}
\definecolor{prdblue}{rgb}{0.133,0.118,0.498}
\usepackage[colorlinks=true, pdfstartview=FitV, linkcolor=prdblue, citecolor= prdblue, urlcolor=prdblue]{hyperref}

\usepackage{CJKutf8}

\usepackage[mathscr,scaled=1.15]{urwchancal}
\DeclareFontFamily{OT1}{pzc}{}
\DeclareFontShape{OT1}{pzc}{m}{it}%
{<-> s * [1.15] pzcmi7t}{}
\DeclareMathAlphabet{\mathpzc}{OT1}{pzc}{m}{it}

\biboptions{sort&compress}

\journal{Physics Letters B}

\begin{document}

\begin{CJK}{UTF8}{song}

\begin{frontmatter}

\title{$\,$\\[-7ex]\hspace*{\fill}{\normalsize{\sf\emph{Preprint no}. NJU-INP 061/22}}\\[1ex]
Fresh look at experimental evidence for odderon exchange
}

\author[NJU,INP]{Zhu-Fang Cui}
\ead{phycui@nju.edu.cn}

\author[ECT]{Daniele Binosi}%
\ead{binosi@ectstar.eu}

\author[NJU,INP]{Craig D. Roberts\corref{cor2}}
\ead{cdroberts@nju.edu.cn}

\author[HZDR,RWTH]{Sebastian M.~Schmidt}%
\ead{s.schmidt@hzdr.de}

\author[ECT]{D.\,N.~Triantafyllopoulos}
\ead{trianta@ectstar.eu}

\address[NJU]{
School of Physics, Nanjing University, Nanjing, Jiangsu 210093, China}
\address[INP]{
Institute for Nonperturbative Physics, Nanjing University, Nanjing, Jiangsu 210093, China}

\address[ECT]{
European Centre for Theoretical Studies in Nuclear Physics
and Related Areas, Villa Tambosi, Strada delle Tabarelle 286, I-38123 Villazzano (TN), Italy}

\address[HZDR]{
Helmholtz-Zentrum Dresden-Rossendorf, Dresden D-01314, Germany}

\address[RWTH]{
RWTH Aachen University, III. Physikalisches Institut B, Aachen D-52074, Germany}


\begin{abstract}
Theory suggests that in high-energy elastic hadron+hadron scattering, $t$-channel exchange of a family of colourless crossing-odd states -- the odderon -- may generate differences between $p\bar p$ and $pp$ cross-sections in the neighbourhood of the diffractive minimum.  Using a mathematical approach based on interpolation via continued fractions enhanced by statistical sampling, we develop robust comparisons between $p\bar p$ elastic differential cross-sections measured at $\surd s=1.96\,$TeV by the D0 Collaboration at the Tevatron and function-form-unbiased extrapolations to this energy of kindred $pp$ measurements at $\surd s /{\rm TeV} = 2.76, 7, 8, 13$ by the TOTEM Collaboration at the LHC and a combination of these data with earlier cross-section measurements at $\surd s/{\rm GeV} = 23.5, 30.7, 44.7, 52.8, 62.5$ made at the internal storage rings.  Focusing on a domain that straddles the diffractive minimum in the $p\bar p$ and $pp$ cross-sections, we find that these two cross-sections differ at the $(2.2-2.6)\sigma$ level; hence, supply evidence with this level of significance for the existence of the odderon.  If combined with evidence obtained through different experiment-theory comparisons, whose significance is reported to lie in the range $(3.4-4.6)\sigma$, one arrives at a $(4.0 -  5.2)\sigma$ signal for the odderon.
\end{abstract}

\begin{keyword}
diffractive production \sep
high-energy hadron+hadron interactions \sep
odderon \sep
Regge phenomenology \sep
Schlessinger point method \sep
strong interactions in the standard model of particle physics
\end{keyword}

\end{frontmatter}

\end{CJK}

\section{Introduction}
\label{Sec1}
It has been known empirically for more than forty years \cite{Baksay:1978sg} that high-energy total and elastic hadron+hadron cross-sections increase slowly with $s$, the square of the total centre-of-mass energy, being bounded above by ${\rm constant} \times \ln^2 s/s_0$ on $s\gg s_0$, where $s_0 \sim m_p^2$ and $m_p$ is the proton mass.  Today, the behaviour is confirmed on a large $s$-domain \cite[Sect.\,20]{Zyla:2020zbs}; and is typically described using Regge phenomenology \cite{Donnachie:2002en, Gribov:2003nw, Ewerz:2003xi}.

In the Regge approach, high-energy hadron+hadron elastic scattering amplitudes receive Regge pole contributions of the form
\begin{equation}
T_{\rm el}(s,t) \propto \eta f(t) (s/s_0)^{\alpha(t)} ,
\end{equation}
where $t$ is the squared four-momentum transfer, $f(t)$ is some reaction-dependent structure factor, $\eta = \pm 1$ is the \emph{signature} of the Regge pole and $\alpha(t)$ is its \emph{trajectory}.
The signature is significant, \emph{e.g}., $\eta = +1$ Regge poles contribute equally to both proton+proton ($pp$) and proton+antiproton ($p\bar p$) scattering whereas $\eta = -1$ poles give opposite sign contributions to these reactions.

Using the optical theorem, which relates the total cross-section to the imaginary part of the elastic amplitude, one finds:
\begin{equation}
s \sigma_{\rm tot} = {\rm Im}T_{\rm el}(s,t=0) \Rightarrow \sigma_{\rm tot} \stackrel{s\gg s_0}{\sim} \sum_R \eta_R (s/s_0)^{\alpha_R(0) - 1},
\end{equation}
where the sum runs over the number of contributing Regge poles.  The Regge pole with the largest intercept, \emph{i.e}., value of $\alpha_R(0)$, dominates $\sigma_{\rm tot}$ at the highest values of $s$.  This pole is called the Pomeron \cite{Gell-Mann:1962daa}, denoted $\mathbb P$.  It is associated with $\eta_{\mathbb P}=+1$ and \cite{Donnachie:1983hf}
\begin{equation}
\alpha_{\mathbb P}(t) = 1.0808 + 0.25t\,.
\end{equation}
Within the context of quantum chromodynamics (QCD), $\mathbb P$ is thought to represent the exchange of a family of colourless crossing-even states, with two-gluon exchange being the simplest contributor \cite{Forshaw:1997dcA}.

A competing or, better, complementary Regge trajectory, known now as the odderon, with $\alpha_{\mathbb O}(0) \simeq 1$ but $\eta_{\mathbb O}=-1$, was also considered as a possible explanation for the $\ln^2 s/s_0$ growth in cross-sections \cite{Lukaszuk:1973nt}.  However, early experiments were unable to unambiguously validate the picture \cite{Erhan:1984mv, Breakstone:1985pe}.  Notwithstanding that, perturbative QCD analyses suggest that the odderon concept is well founded, being linked to the exchange of a family of colourless crossing-odd states, with three-gluon exchange as the leading term and a much smaller coupling to nucleons than exchanges in the Pomeron family \cite{Fukugita:1979mxf, Janik:1998xj, Bartels:1999yt, Ewerz:2003xi}.

Dominance of $\mathbb P$ exchange in total cross-sections and small-angle elastic scattering might explain why the odderon remains hidden.  Yet, there are kinematic domains on which $\mathbb P$ exchange is suppressed, \emph{e.g}., in the neighbourhood of the diffractive minimum in hadron+hadron scattering.  Focusing on this region, one may reasonably expect to find a signal for the odderon in differences between $pp$ and $p\bar p$ elastic scattering.   Exploiting such opportunities, Refs.\,\cite[TOTEM]{TOTEM:2017sdy} and \cite{Csorgo:2019ewn} have argued there is good evidence for an odderon contribution to hadron+hadron scattering amplitudes.

Reviewing these reports, Ref.\,\cite{TOTEM:2020zzr} claims to deliver a comparison of $\surd s=1.96\,$TeV $p \bar p$ elastic differential cross-sections measured at the Fermilab Tevatron \cite[D0]{D0:2012erd} with ``model-independent'' extrapolations to this lower Tevatron energy of $\surd s/{\rm TeV}= 2.76, 7, 8, 13$ $pp$ elastic differential cross-section measurements at the large hadron collider (LHC) \cite[TOTEM]{TOTEM:2018psk, TOTEM:2011vxg, TOTEM:2015oop, TOTEM:2018hki}; and therewith provide a direct demonstration of the odderon's existence.  However, as remarked in Ref.\,\cite[Note\,56]{TOTEM:2020zzr}, there are actually two possible sources of model dependence in the extrapolation:
($\mathpzc a$) the forms chosen for fitting the measured TOTEM LHC $pp$ $(t,d\sigma/dt)$ values as functions of $\surd s$ for subsequent use in extrapolating to the D0 energy;
and ($\mathpzc b$) the function chosen in fitting the TOTEM $pp$ cross-sections to enable interpolation to the $t$ values at which D0 $p\bar p$ measurements were made.
Using a recently refined mathematical approach, \emph{viz}.\ the Schlessinger Point Method (SPM) \cite{PhysRev.167.1411, Schlessinger:1966zz, Tripolt:2016cya, Chen:2018nsg}, which has been widely applied with success in the interpolation and extrapolation of experimental data and theoretical predictions \cite{Cui:2022fyr}, it is possible to eliminate both ($\mathpzc a$) and ($\mathpzc b$).
Herein, therefore, we employ the SPM to revisit and extend the TOTEM analysis and thereby deliver objective results concerning the odderon that are free of function-form-bias.

\section{Analysis of LHC $pp$ data -- Method A}
\label{MethodA}
Working with the TOTEM $pp$ elastic differential cross-sections at $\surd s/{\rm TeV} = 2.76, 7, 8, 13$ \cite{TOTEM:2018psk, TOTEM:2011vxg, TOTEM:2015oop, TOTEM:2018hki}, sketched as $d\sigma/dt$ \emph{vs}.\ $|t|$ in Ref.\,\cite[Fig.\,1]{TOTEM:2020zzr}, we proceed to employ the SPM in developing interpolations and subsequently extrapolations.  The theory underlying the SPM is explained elsewhere \cite[Sect.\,3]{Cui:2022fyr}.  Here, therefore, we only include a few remarks.

{\sf\small SPM background}.
The SPM circumvents any need for a specific choice of fitting function in analysing data.  Honed in numerous applications, especially those which require interpolation and reliable extrapolation, \emph{e.g}., Refs.\,\cite{Binosi:2018rht, Binosi:2019ecz, Eichmann:2019dts, Yao:2021pdy, Cui:2021gzg, Cui:2022fyr}, the SPM builds function-form-unbiased continued-fraction interpolations of data as the foundation for well-constrained extrapolations with quantified uncertainties.

The efficacy of the SPM is founded on its mathematical connection with the Pad\'e approximant; in fact, the procedure might also be called the multipoint Pad\'e approximants technique.  Owing to the special features of analytic functions, then with $N>0$ values of such a function, ${\mathpzc f}$, at discrete real points, ${\mathpzc D}=\{x_i\,|\,i=1,\ldots,N\}$, the SPM is guaranteed to return an accurate representation of ${\mathpzc f}(x)$ within a radius of convergence determined by that one of the function's branch-points which lies nearest to the domain ${\mathpzc D}$.  An elementary example is provided by a monopole: ${\mathpzc f}(x)=1/(1+x)$, represented by function values at $N$ points, with $N$ large.  Using any one of those points, the SPM will precisely reproduce ${\mathpzc f}(x)$.

To counter the problem that perfect function values are practically impossible to obtain, a powerful statistical aspect is introduced.  Namely, one chooses $M \lesssim N/2$ points at random from ${\mathpzc D}$ and works with the associated function values.  Continuing with our elementary example, then using the SPM with each pair, $(x_i,{\mathpzc f}(x_i))$, $i=1,\ldots, M$, one produces a collection of analytic approximations to the monopole whose spread measures the uncertainty inherent in the ``measured'' function values.  Each one of the approximations is of equal quality to the best least-squares fit to ${\mathpzc f}$ as defined by the points in ${\mathpzc D}$.  These ideas extend straightforwardly to more sophisticated functions.  (Practical illustrations are provided elsewhere \cite{DanieleBinosiPresentation}.)

{\small\sf Stage 1}.
Regarding the TOTEM $pp$ elastic differential cross-section measurements,
we consider each value of $\surd s$ separately; and in each case generate $n_R=1\,000$ replica cross-sections according to a binormal distribution with mean equal to the central value of $(|t|,d\sigma/dt)$ and variances equal to the errors in $|t|$, $d\sigma/dt$.  (In the absence of available information to the contrary, we treated those errors as uncorrelated.)
%
Selecting a particular replica, we choose $M$ elements at random, with $M\in {\cal S}_M=\{9,10,11,12,13\}$, and mathematically construct a continued-fraction interpolation based on these $M$ points.
If that interpolator is smooth on $0.1<|t|/{\rm GeV}^2<1$ and, with increasing $|t|$,  displays a diffractive minimum followed by a maximum, then it is retained; otherwise, it is discarded.
This process is repeated until $n_{\mathpzc I}=1\,000$ independent ``physical'' interpolators are obtained for the given value of $M$.
It is then repeated for a new value of $M$, and so on.
In this way, we arrive at $5\,000\,000$ independent interpolators for each value of $\surd s$.

{\small\sf Stage 2}.
In the next step, we emulate and extend the notion explained in connection with Ref.\,\cite[Fig.\,3a]{TOTEM:2020zzr}.
Namely, motivated by the shape of the measured $pp$ elastic differential cross-section, we define a set of 29 characteristic points within the measured $t$ domain:
$8$ are those chosen in Ref.\,\cite[Fig.\,3a]{TOTEM:2020zzr};
a ``mid3'' point is introduced, being the smallest-$|t|$ intercept of the measured cross-section with the straight line passing through the TOTEM ``mid1'' and ``mid2'' points;
%
points are added at ``{\rm bump}+2.5'' and ``{\rm bump}+7.5'';
and an additional $18=6\times 3$ points are included by sampling in steps of $A/8$, instead of $A/2$, where $A$ is the ``bump'' height minus ``dip-2'' minimum, giving $6$ ``bump2''-to-``dip2'' interior points, and keeping all three cross-section intersections.
This $29/8$-fold expansion of the cross-section characterising set is enabled by the huge number of interpolators we have constructed; and it equips us to develop a more robust SPM extrapolation of the $pp$ cross-section at the $\surd s = 1.96\,$TeV D0 $|t|$ bins.

\begin{figure*}[!t]
\vspace*{0ex}

\leftline{\hspace*{0.5em}{\large{\textsf{A}}}}
\vspace*{-1.5ex}

\includegraphics[width=0.99\textwidth]{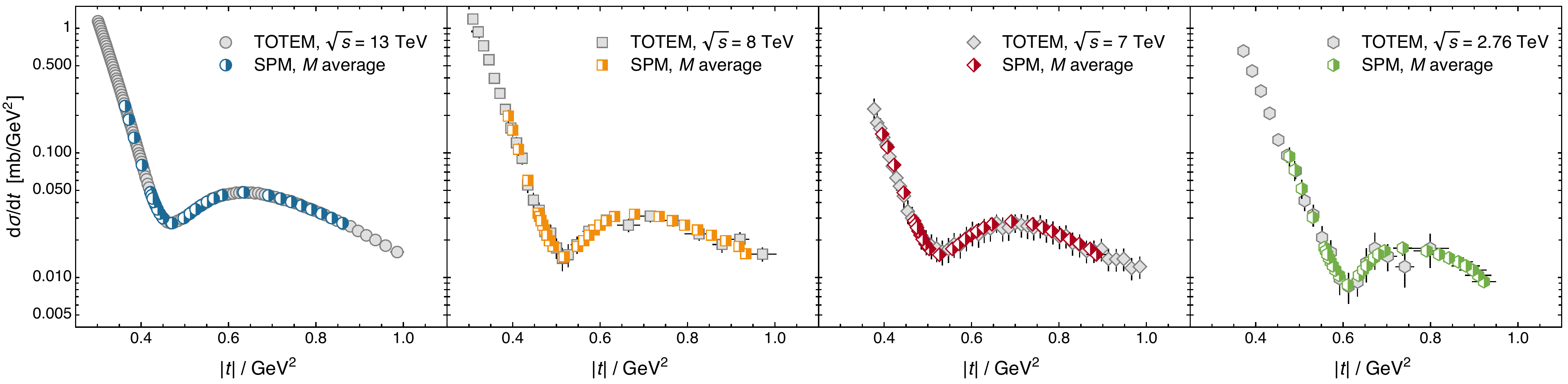}
\vspace*{+1ex}

\leftline{\hspace*{0.5em}{\large{\textsf{B}}}}
\vspace*{-1.5ex}

\includegraphics[width=0.99\textwidth]{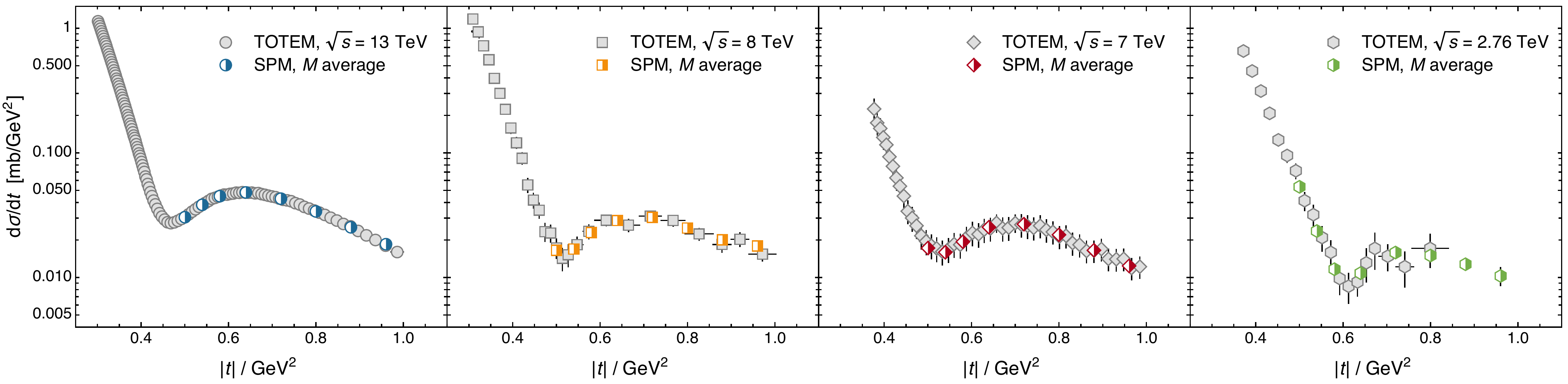}
\vspace*{+1ex}

\leftline{\hspace*{0.5em}{\large{\textsf{C}}}}
\vspace*{-1.5ex}

\includegraphics[width=0.99\textwidth]{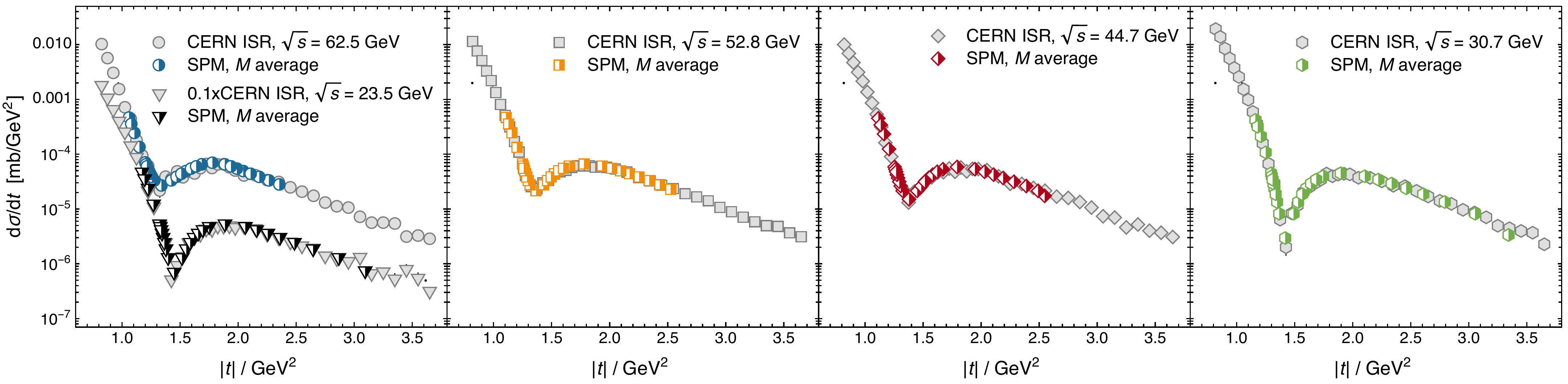}
\vspace*{+1ex}

\leftline{\hspace*{0.5em}{\large{\textsf{D}}}}
\vspace*{-1.5ex}
\includegraphics[width=0.99\textwidth]{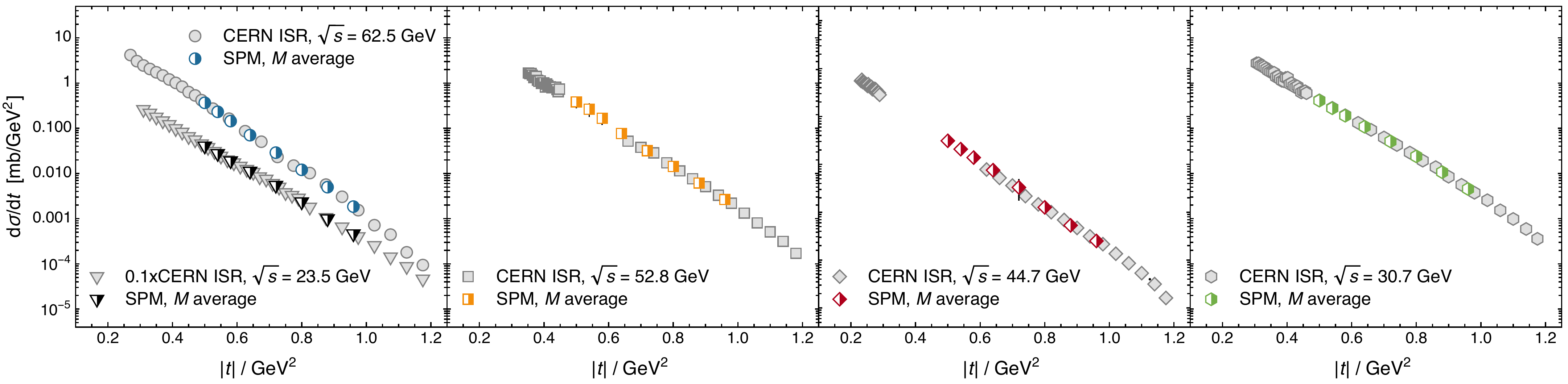}
\caption{\label{F1TOTEM}
\emph{First row}\,--\,{\sf A}.
LHC measurements of $pp$ elastic differential cross-sections at
$\surd s/{\rm TeV} = 2.76, 7, 8, 13$ \cite{TOTEM:2018psk, TOTEM:2011vxg, TOTEM:2015oop, TOTEM:2018hki} -- grey-shaded shapes.
Values obtained using the SPM to determine the associated cross-sections at each one of the 29 characteristic points defined in Sect.\,\ref{MethodA}\,--\,{\sf\footnotesize Stage 2} -- half-shaded coloured shapes.
\emph{Second row}\,--\,{\sf B}.  As in Row~{\sf A} except SPM used to determine the associated cross-sections at each one of the 8 D0 bins in Eq.\,\eqref{D0bins}.
\emph{Third row}\,--\,{\sf C}.  As in Row~{\sf A} except the SPM was applied to ISR measurements of $pp$ elastic differential cross-sections at $\surd s ({\rm GeV}) = 23.5, 30.7, 44.7, 52.8, 62.5$ \cite{Amaldi:1979kd} (grey-shaded shapes) to determine the associated cross-sections at each of the 29 characteristic points (half-shaded coloured shapes).
\emph{Fourth row}\,--\,{\sf D}.  As in Row~{\sf B} except the SPM was applied to ISR measurements of $pp$ differential cross-sections, which are relatively featureless on the $|t|$-domain sampled by D0.
}
\end{figure*}

Working with a given value of $\surd s$ and a particular choice for $M$, the value of the cross-section at any one of the characteristic points, $|t_c|$, is
\begin{equation}
\label{average}
\overline{\frac{d\sigma}{dt}}_{|t_c|,M}
=\frac{1}{n_{\mathpzc I}} \sum_{j=1}^{n_{\mathpzc I}} {\mathpzc I}_j^M(|t_c|)
\end{equation}
with uncertainty, $\epsilon_{|t_c|,M}$, given by the associated standard deviation
\begin{equation}
\label{averageSD}
\epsilon^2_{|t_c|,M} = \frac{1}{n_{\mathpzc I}}
\sum_{j=1}^{n_{\mathpzc I}}\left[ {\mathpzc I}_j^M(|t_c|) - \overline{\frac{d\sigma}{dt}}_{|t_c|,M}\right]^2 .
\end{equation}
Repeating for every value of $M \in {\cal S}_M$, then one has a set of five results for the cross-section at this characteristic point:
$\{(\overline{d\sigma/dt}_{|t_c|,M},\epsilon_{|t_c|,M})\,|\, M\in {\cal S}_M\}$;
and the final value of the $\surd s$ cross-section at $|t_c|$ is the uncertainty weighted average obtained from this set.
The above procedure is repeated at every one of the 29 characteristic points and for each value of $\surd s /{\rm TeV} = 2.76, 7, 8, 13$; and this yields the results displayed in Fig.\,\ref{F1TOTEM}A.

\begin{figure}[t]
\vspace*{0ex}

\leftline{\hspace*{0.5em}{\large{\textsf{A}}}}
\vspace*{-1ex}

\includegraphics[width=0.4\textwidth]{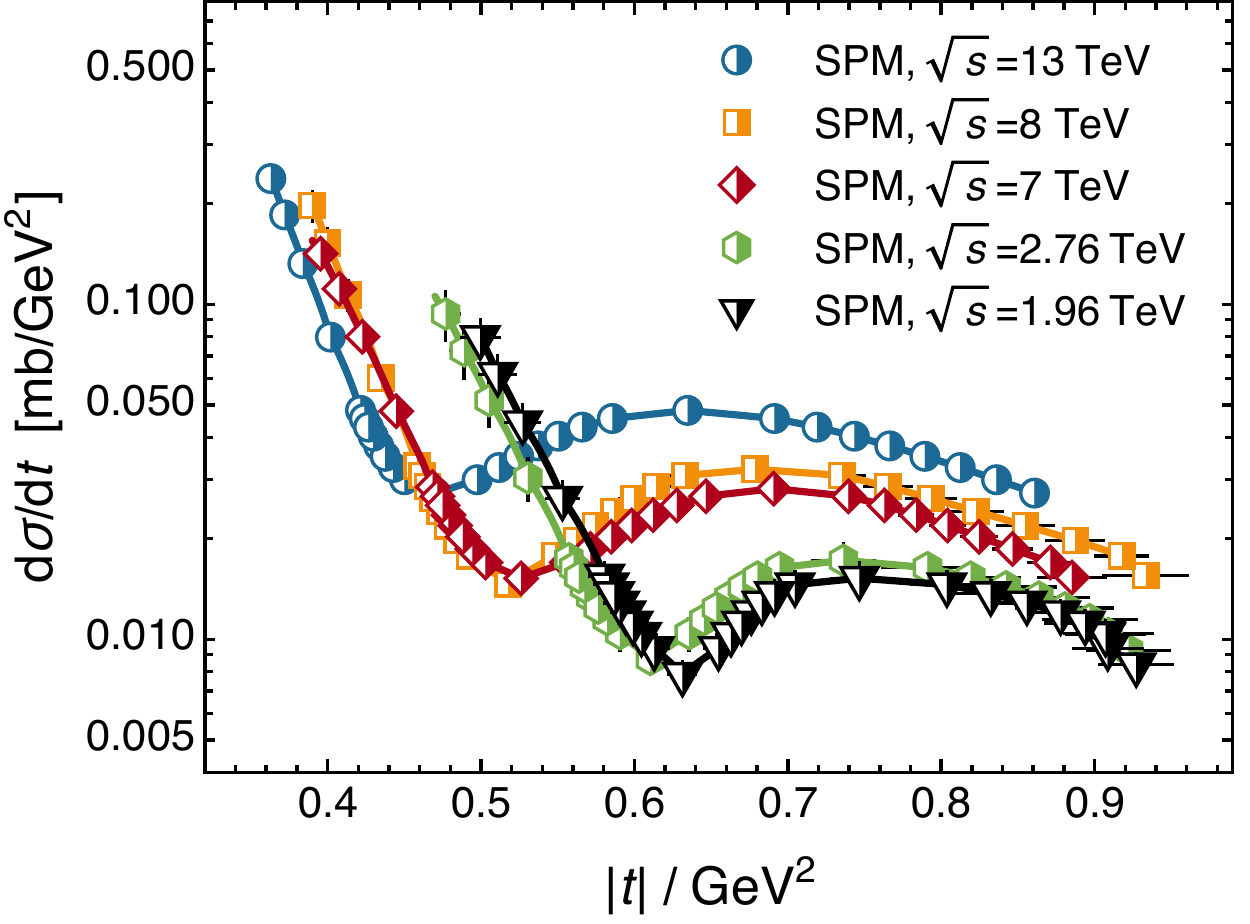}
\vspace*{+2ex}

\leftline{\hspace*{0.5em}{\large{\textsf{B}}}}
\vspace*{-1ex}
\includegraphics[width=0.4\textwidth]{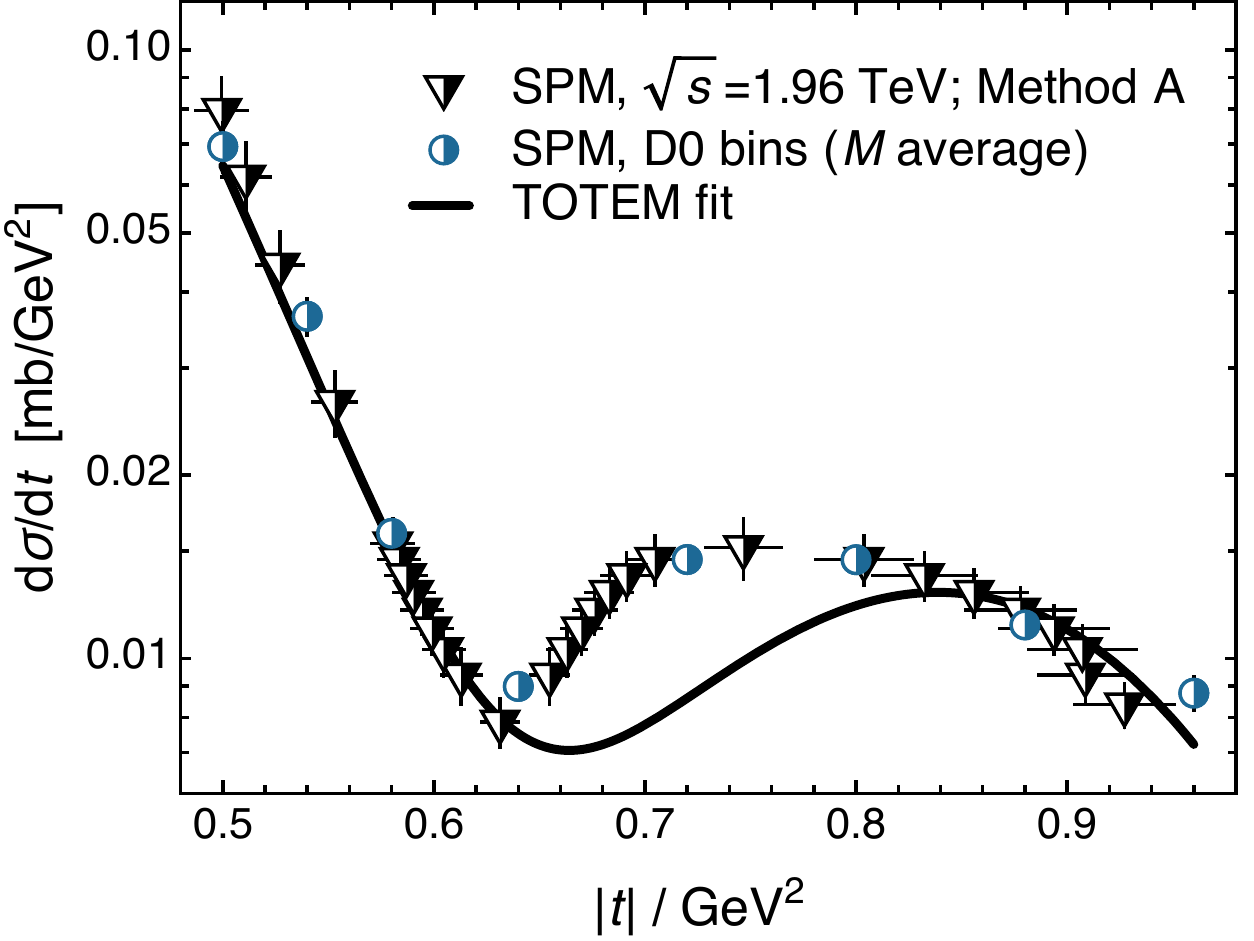}

%
\caption{\label{F2TOTEM}
\emph{Top panel}\,--\,{\sf A}.
SPM interpolations of TOTEM LHC measurements of $pp$ elastic differential cross-section \cite{TOTEM:2018psk, TOTEM:2011vxg, TOTEM:2015oop, TOTEM:2018hki}, highlighting values at the characteristic points defined in the text preceding Eq.\,\eqref{average}, compared with the resulting SPM extrapolation to $\surd s = 1.96\,$TeV (half-shaded down-triangles).
\emph{Bottom panel}\,--\,{\sf B}.  TOTEM measurements of $pp$ elastic differential cross-section extrapolated to $\surd s = 1.96\,$GeV: half-shaded down-triangles -- SPM using characteristic points; and half-shaded circles -- SPM when working with the D0 $|t|$ bins, Eq.\,\eqref{D0bins}.  Solid curve -- function-form-dependent extrapolation in Ref.\,\cite{TOTEM:2020zzr}.
}
\end{figure}

{\small\sf Stage 3}.
The SPM results displayed in Fig.\,\ref{F1TOTEM}A define four-element sets that encode the $s$-dependence of the $pp$ elastic differential cross-section at each one of the 29 characteristic points.  This information must now be used to extrapolate the cross-sections to the D0 energy, $\surd s = 1.96\,$TeV.
Proceeding as follows, the SPM can again be used to accomplish this.
(\emph{i}) At each $|t_c|$, generate $n_R=1\,000$ replicas of the cross-section value using a normal distribution with mean equal to the function value and variance identified with its uncertainty.
(\emph{ii}) Working with $M\in {\cal S}^\prime_M=\{4,3,2\}$, then for each $M$ construct $1\,000$ continued-fraction interpolations, requiring only that they be smooth on $0<\surd s/{\rm TeV}<15$.
(\emph{iii}) For each $M\in {\cal S}^\prime_M$, obtain the average and uncertainty in analogy with Eqs.\,\eqref{average}, \eqref{averageSD}.
(\emph{iv}) Define the final value of the $\surd s = 1.96\,$TeV $pp$ cross-section at $|t_c|$ as the uncertainty-weighted average computed from this set.
In this way, one obtains the results displayed in Fig.\ref{F2TOTEM}.

Regarding Fig.\ref{F2TOTEM}A, it is evident that the SPM extrapolation of the $pp$ elastic differential cross-sections to $\surd s=1.96\,$TeV is smooth, \emph{e.g}., the difference between the $\surd s/{\rm TeV} = 2.76, 1.96$ curves matches well with that between the measured $\surd s/{\rm TeV} = 8, 7$ cross-sections.
This presents a marked contrast with the results depicted in Ref.\,\cite[Fig.\,1]{TOTEM:2020zzr}, where the difference between the measured $\surd s= 2.76\,$TeV cross-section and the function-form-dependent extrapolation to $\surd s = 1.96\,$TeV is significantly bigger than that between the measured $\surd s/{\rm TeV} = 8, 7$ cross-sections, especially on a large neighbourhood within the diffractive minimum.
Turning to Fig.\ref{F2TOTEM}B, one sees, as could be anticipated from the observations just made, that the extrapolation to $\surd s=1.96\,$TeV made using the SPM, which eliminates bias associated with the choices of functions for fitting and extrapolation, differs materially from that in Ref.\,\cite{TOTEM:2020zzr} just where discrepancies are most important, \emph{i.e}., within the diffraction minimum.

\begin{figure*}[!t]
\includegraphics[width=1\textwidth]{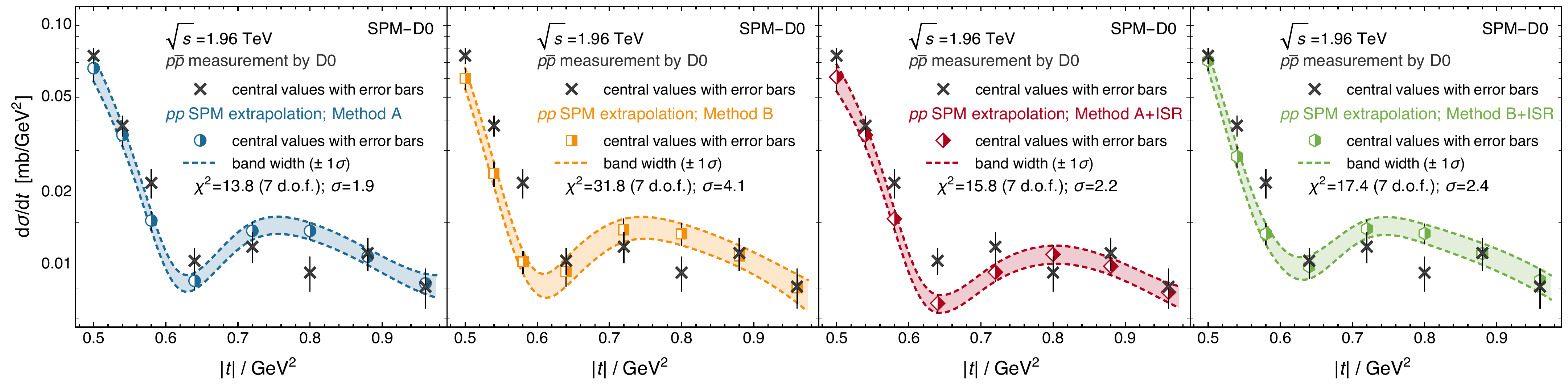}
\caption{\label{F3TOTEM}
Crosses, each panel -- $p\bar p$ elastic differential cross-sections measured by D0 \cite{D0:2012erd}.
The panels are identified from left-to-right as Images 1 -- 4.
\emph{Image~1}.  Method A comparison -- 4 stage process, Sect.\,\ref{MethodA}.
Half-shaded circles -- SPM extrapolation of TOTEM $pp$ elastic differential cross-sections.
\emph{Image~2}.  Method B -- 3 stage process, Sect.\,\ref{MethodB}.
Half-shaded squares  -- SPM extrapolation of LHC $pp$ elastic differential cross-sections.
\emph{Image~3}.  Method A$_{+{\rm ISR}}$ comparison -- 4 stage process, Sect.\,\ref{MethodAISR}.
Half-shaded diamonds -- SPM interpolation of ISR and TOTEM $pp$ elastic differential cross-sections.
\emph{Image~4}.  Method B$_{+{\rm ISR}}$ comparison -- 3 stage process, Sect.\,\ref{MethodBISR}.
Half-shaded hexagons -- SPM interpolation of ISR and TOTEM $pp$ elastic differential cross-sections.
In each panel: the SPM cross-sections are rescaled by $0.954 \pm 0.071$ for the reasons explained in Sect.\,\ref{MethodA}\,--\,{\footnotesize\sf Stage 4}; and the $\pm 1\sigma$ uncertainty on the SPM results are marked by the associated shaded bands.
}
\end{figure*}

At this point, we have introduced the SPM, validated the approach in comparisons with real LHC data -- Figs.\,\ref{F1TOTEM}A, \ref{F2TOTEM}A, and highlighted a potential issue with the function-form-dependent interpolations and extrapolations in Ref.\,\cite{TOTEM:2020zzr}\,-\,Fig.\,\ref{F2TOTEM}B.  The next step is to repeat the SPM analysis, focusing not on the characteristic points identified in Sect.\,\ref{MethodA}\,-\,{\small\sf Stage 2} but on the $|t|$-bin values at which D0 measurements were actually made \cite{D0:2012erd}:
\begin{equation}
\label{D0bins}
|t|/{\rm GeV}^2 = \{0.50, 0.54, 0.58, 0.64, 0.72, 0.80, 0.88, 0.96\}\,.
\end{equation}
The result is also displayed in Fig.\,\ref{F2TOTEM}B: evidently, the two procedures yield compatible results.
%

{\small\sf Stage 4}.
The final step in building a fair comparison between the $p\bar p$ elastic differential cross-sections measured by D0 \cite{D0:2012erd} and the SPM extrapolation of the TOTEM $pp$ cross-sections is rescaling of the latter by a factor of $0.954 \pm 0.071$ so that their values at the optical point, $t=0$, are the same as that for $p\bar p$.  This is the procedure employed in Ref.\,\cite{TOTEM:2020zzr}.  It would be the case if only crossing-even exchanges were involved in both reactions; and it leads us to the comparison depicted in Fig.\,\ref{F3TOTEM}\,-\,Image~1.

Working with the results in Fig.\,\ref{F3TOTEM}\,-\,Image~1, we employ a $\chi^2$ test to determine the probability that the D0 and TOTEM cross-sections agree.  Namely,
(\emph{i}) we subtract the two cross-sections and sum their errors in quadrature;
(\emph{ii}) calculate the $\chi^2$ sum with reference to the null hypothesis;
and (\emph{iii}) compute the $p$-value and corresponding $\sigma$ significance.
With $7$ degrees-of-freedom, because we have imposed a normalisation condition, one thereby obtains a $p$-value of 5.4\%, meaning that the D0 and TOTEM cross-sections disagree at a level of
\begin{equation}
{\rm S}^{\rm A}_{\mathbb O}=  1.9\sigma\,.
\end{equation}
Alone, this is insufficient to support a confident claim for odderon discovery.
Our result should be contrasted with that estimated in Ref.\,\cite{TOTEM:2020zzr} using function-form-dependent interpolations and extrapolations: $p=0.061$\% $\Rightarrow 3.4\sigma$.
The disagreement owes to the mismatch on $0.64\leq |t|/{\rm GeV}^2\leq 0.84$, evident in Fig.\,\ref{F2TOTEM}B, between the Ref.\,\cite{TOTEM:2020zzr} extrapolation and the function-form-unbiased SPM result.

\section{Analysis of LHC $pp$ data -- Method B}
\label{MethodB}
Familiarity with the SPM reveals that there is an obvious alternative to Method A in extrapolating the TOTEM LHC $pp$ cross-sections.  Namely:
the SPM can be used to construct a large number of interpolators for each of the TOTEM cross-sections;
those interpolators can be employed directly to deliver results at the D0 $|t|$ bins, Eq.\,\eqref{D0bins};
and the SPM can then be used again to extrapolate those results to the D0 energy, $\surd s = 1.96\,$TeV.

We implement Method B as follows.
Each value of $\surd s/{\rm TeV} = 2.76,7,8,13$ is considered separately; and in each case, $n_R=1\,000$ replica cross-sections are generated according to a normal distribution with mean equal to the central value of the cross-section and variance equal to its error.  (In this case, the $|t|$ value is fixed with no uncertainty.)
Selecting a particular replica, we choose $M$ elements at random, with $M\in {\cal S}_M$, and mathematically compute a continued-fraction interpolation based on these $M$ points.
Again, if that interpolator is a smooth function on $0.1<|t|/{\rm GeV}^2<1$ and, with increasing $|t|$,  displays a diffractive minimum followed by a maximum, then it is retained; otherwise, it is discarded.
This process is repeated until $n_{\mathpzc I}=1\,000$ independent ``physical'' interpolators are obtained for the given value of $M$.
It is then repeated for a new value of $M$, and so on.
In this way, we arrive at $5\,000\,000$ independent interpolators for each value of $\surd s$, from which we obtain values for all $pp$ elastic differential cross-sections with quantified uncertainties at each of the D0 $|t|$ bins, as displayed in Fig.\,\ref{F1TOTEM}B.

The SPM results displayed in Fig.\,\ref{F1TOTEM}B define four-element sets that encode the $s$-dependence of the $pp$ elastic differential cross-section at each one of the 8 D0 $|t|$ bins.
Following the procedure employed in Sect.\,\ref{MethodA}\,--\,{\small\sf Stage 3}, the SPM can now be used again to extrapolate this information to the D0 energy, $\surd s = 1.96\,$TeV.  Finally, rescaling those results, as described in Sect.\,\ref{MethodA}\,--\,{\small\sf Stage 4}, we arrive at the comparison drawn in Fig.\,\ref{F3TOTEM}\,-\,Image~2.
In this case, using the same $\chi^2$ test employed in connection with Fig.\,\ref{F3TOTEM}\,-\,Image~1, the value $p=0.0044$\% is obtained; thus, when compared using Method B, the level of disagreement between D0 and extrapolated TOTEM cross-sections is
\begin{equation}
{\rm S}^{\rm B}_{\mathbb O}=  4.1\sigma\,.
\end{equation}

\section{Including ISR data  -- Method A}
\label{MethodAISR}
Thus far we have only worked with LHC measurements, extrapolating down to the D0 energy.  However, a substantial amount of $pp$ elastic scattering data was collected at the Intersecting Storage Rings (ISR) roughly forty-five years ago \cite{Amaldi:1979kd} at $\surd s/{\rm GeV} = 23.5, 30.7, 44.7, 52.8, 62.5$.  The energies involved are two orders-of-magnitude smaller than those in the D0 and LHC experiments, but this alone is no impediment to their inclusion in our analysis.  The SPM is equally sound whether being used to extrapolate down or up in energy.  Furthermore, after including the lower-energy ISR data, one then has constraints at both ends of the energy spectrum, in which case the SPM is actually providing a constrained interpolation instead of an extrapolation.

We now focus, therefore, on the ISR data \cite{Amaldi:1979kd}.  They differ from the TOTEM data in not reporting an uncertainty on the $|t|$ values; hence, when employing Sect.\,\ref{MethodA}\,--\,{\small\sf Stage 1}, we generate $n_R=1\,000$ replica cross-sections according to a normal distribution with mean equal to the central value of $d\sigma/dt$ and variance equal to the associated error.  Otherwise, the procedure is practically identical.  We work with $M\in {\cal S}_M$, as before, but require that accepted interpolators are smooth functions on the expanded domain $0.1<|t|/{\rm GeV}^2<4$, still displaying a diffractive minimum followed by a maximum with increasing $|t|$.  The domain extension is required because the diffraction minimum shifts to a larger value of $|t|$ as $\surd s$ is reduced.

The steps described in Sect.\,\ref{MethodA}\,--\,{\small\sf Stage 2} are repeated without modification and yield the results depicted in Fig.\,\ref{F1TOTEM}C.

{\small\sf Stage 3}$^\prime$.
At this point we have at our disposal the SPM results in Figs.\,\ref{F1TOTEM}A, \ref{F1TOTEM}C, which define nine-element sets -- 4 TOTEM sets above the D0 energy and 5 ISR sets below -- that encode the $s$-dependence of the $pp$ elastic differential cross-section at each one of the 29 characteristic points.  This information can now be used to interpolate (instead of extrapolate) the cross-sections to the D0 energy.  Again, the SPM is used to accomplish this.
(\emph{i}) At each $|t_c|$, generate $n_R=1\,000$ replicas of the cross-section value using a normal distribution with mean equal to the function value and variance identified with its uncertainty.
(\emph{ii}) Working with $M\in {\cal S}^{\prime\prime}_M=\{5,6,7,8,9\}$, then for each $M$ construct $1\,000$ continued-fraction interpolations, requiring only that they be smooth on $0<\surd s/{\rm TeV}<15$.
(\emph{iii}) For each $M\in {\cal S}^{\prime\prime}_M$, obtain the average and uncertainty in analogy with Eqs.\,\eqref{average}, \eqref{averageSD}.
(\emph{iv}) Define the final value of the $\surd s = 1.96\,$TeV $pp$ cross-section at $|t_c|$ as the uncertainty-weighted average computed from this set.

Working with the cross-sections thus obtained, we implement the Sect.\,\ref{MethodA}\,--\,{\small\sf Stage 4} rescaling and arrive at the comparison drawn in Fig.\,\ref{F3TOTEM}\,-\,Image~3.
Now using the $\chi^2$ test, one finds $p=2.7$\%, which means that when compared using Method A, the D0 and interpolated ISR and TOTEM cross-sections disagree with significance
\begin{equation}
{\rm S}^{{\rm A}_{+{\rm ISR}}}_{\mathbb O}=  2.2\sigma\,.
\end{equation}

\section{Including ISR data  -- Method B}
\label{MethodBISR}
One can also utilise the ISR data to expand the Sect.\,\ref{MethodB} scheme, \emph{i.e}.,
first use the SPM to construct a large number of interpolators for each of the ISR cross-sections;
then employ these interpolators to obtain results directly at the D0 $|t|$ bins.
Since all D0 $|t|$-bins lie below $|t|=1\,$GeV$^2$, then one can apply Method B to the ISR data precisely as described in Sect.\,\ref{MethodB} and this yields the results displayed in Fig.\,\ref{F1TOTEM}D, the panels of which compare the original data with the SPM interpolations to each of the D0 $|t|$ bins.

\begin{figure}[!t]
\includegraphics[width=0.4\textwidth]{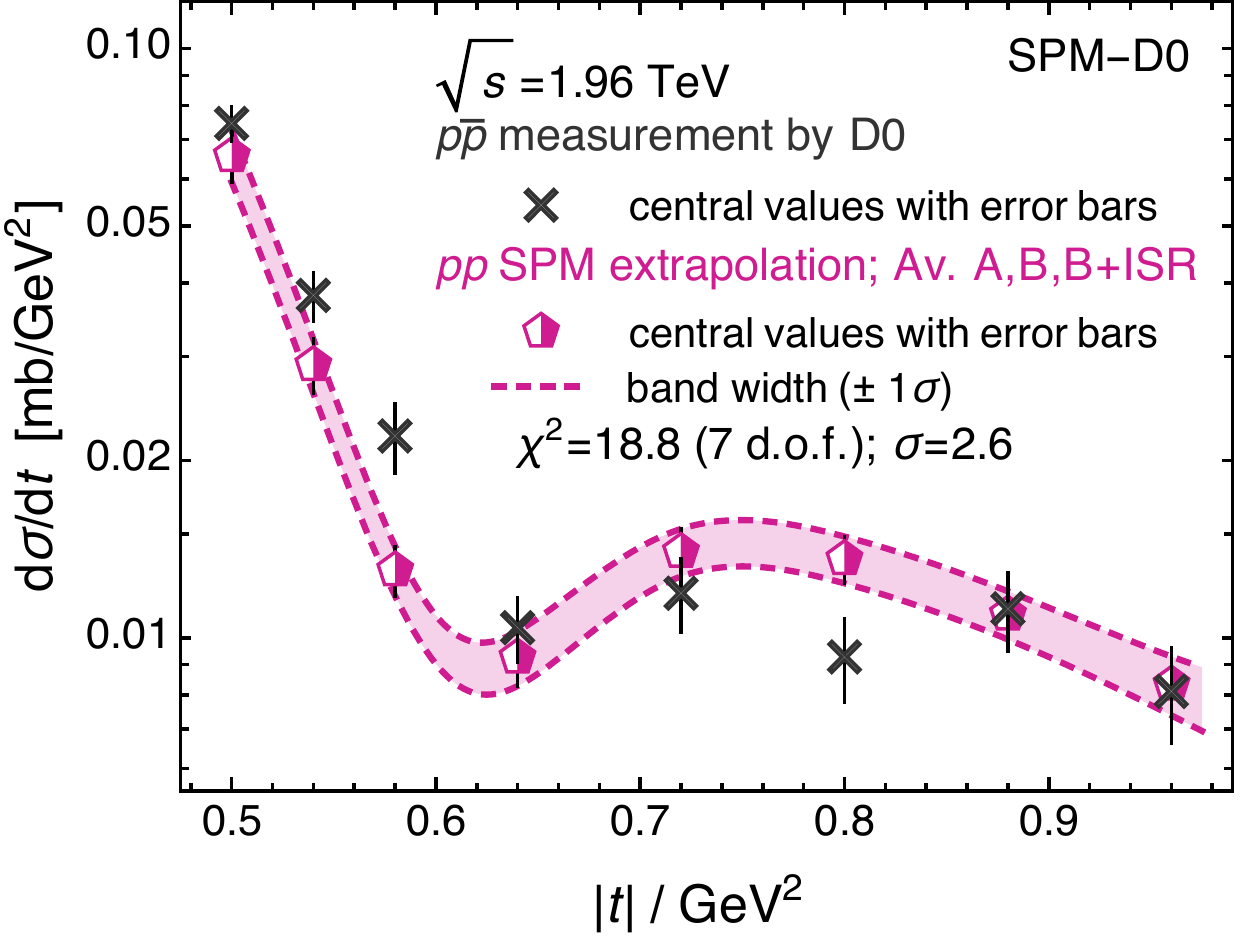}
\caption{\label{F4TOTEM}
Pentagons -- average of the SPM results obtained as described in Sects.\,\ref{MethodA}, \ref{MethodB}, \ref{MethodBISR}.  The $\pm 1\sigma$ uncertainty on this final SPM result is highlighted by the associated shaded band.
Black ``{\sf x}'' marks -- $p\bar p$ elastic differential cross-sections measured by D0 \cite{D0:2012erd}.
}
\end{figure}

As in Sect.\,\ref{MethodAISR}\,-\,{\small\sf Stage 3}$^\prime$, by combining the SPM results in Fig.\,\ref{F1TOTEM}D with those in Fig.\,\ref{F1TOTEM}B, we arrive at eight nine-element sets, each one of which encodes the $s$-dependence of the $pp$ elastic differential cross-section at a given D0 $|t|$ bin.  Here, again, the SPM can be used to interpolate this information so that D0 $|t|$-bin cross-section values are obtained at the D0 energy.  Rescaling those results, as described in Sect.\,\ref{MethodA}\,--\,{\small\sf Stage 4}, we arrive at the comparison drawn in Fig.\,\ref{F3TOTEM}\,-\,Image~4.
Using the $\chi^2$ test employed in connection with Fig.\,\ref{F3TOTEM}\,-\,Image~1, the value $p=1.5$\% is obtained.  Consequently, when compared using Method B, the D0 and interpolated ISR and TOTEM cross-sections disagree at a level of
\begin{equation}
{\rm S}^{{\rm B}_{+{\rm ISR}}}_{\mathbb O}=  2.4\sigma\,.
\end{equation}

\section{Combining compatible $pp$ cross-sections}
\label{MethodABBLE}
To the eye, the SPM results in Fig.\,\ref{F3TOTEM}\,-\,Images~1, 2, 4 are compatible, but that in Image~3 is different from the others.  This observation can be quantified by using the following $\chi^2$ test.
(\emph{i}) Select any two SPM cross-section at the D0 bins and sum their errors (because the two extrapolations are not truly independent);
(\emph{ii}) calculate the $\chi^2$ sum with reference to the null hypothesis;
and (\emph{iii}) compute the $p$-value and corresponding $\sigma$ significance.
One finds in this way that a pairwise comparison between any two of Images~1, 2, 4 in Fig.\,\ref{F3TOTEM} returns a $<1\sigma$ difference.  On the other hand, the result in Image~3 differs from those in Images~1, 2, 4 by $1.3\sigma$, $2.0\sigma$, $1.8\sigma$, respectively.
%

Having mathematically established compatibility between the three SPM extrapolations in Fig.\,\ref{F3TOTEM}\,-\,Images~1, 2, 4, one is justified in averaging them to obtain the final result, which is displayed in Fig.\,\ref{F4TOTEM}.
Working with this final result, using the same $\chi^2$ test employed in connection with Figs.\,\ref{F3TOTEM}, one obtains $p=0.9$\%; hence, the level of disagreement between D0 and SPM analyses of ISR and TOTEM cross-sections is
\begin{equation}
\label{Sigodderon}
{\rm S}^{\rm A+B+B_{+{\rm ISR}}}_{\mathbb O} = 2.6\sigma\,.
\end{equation}

\begin{figure}[!t]
\includegraphics[width=0.4\textwidth]{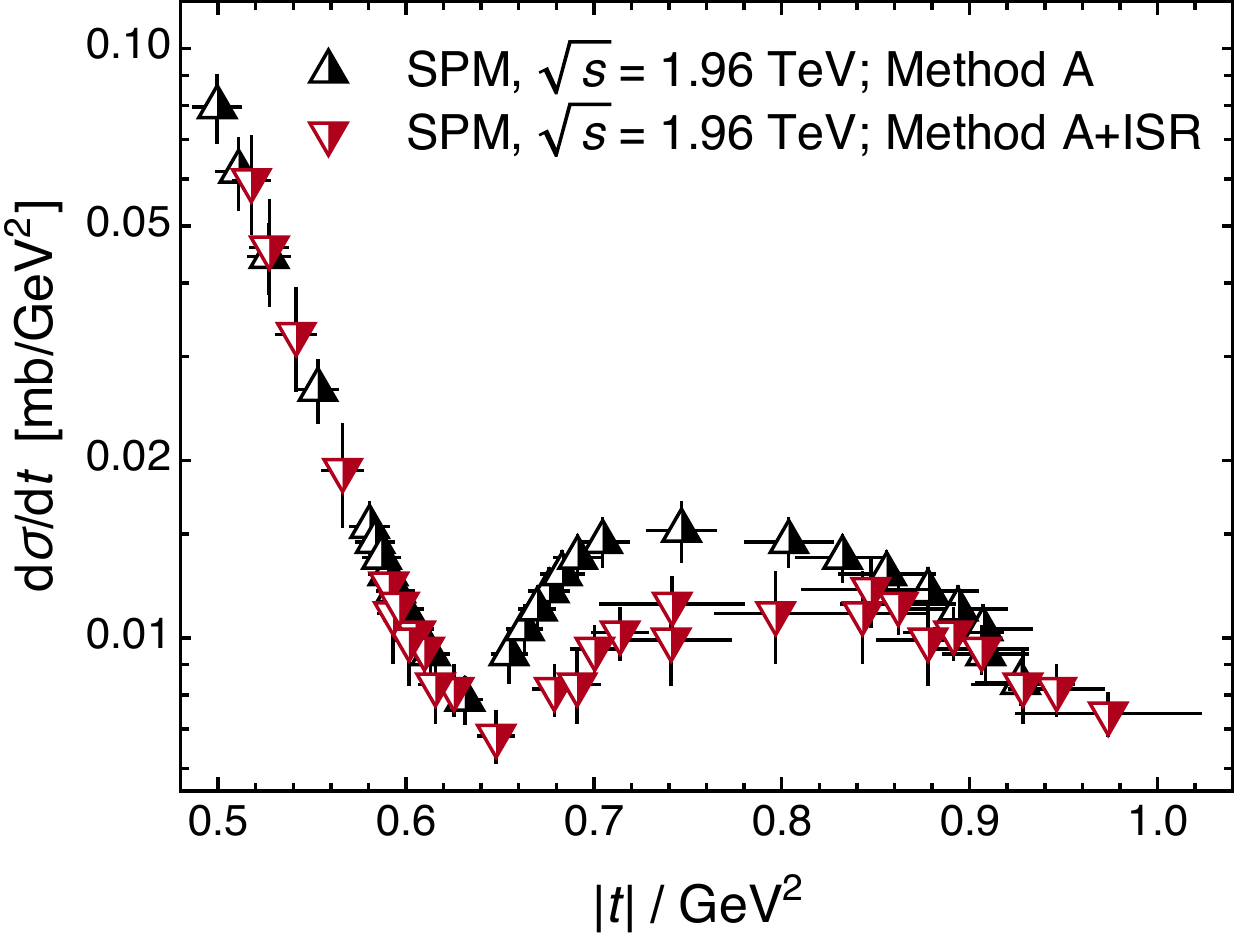}
\caption{\label{F5TOTEM}
Comparison between Method A extrapolation of TOTEM data, described in Sect.\,\ref{MethodA} --
up-triangles, and SPM interpolation of ISR and TOTEM data, discussed in Sect.\,\ref{MethodAISR} --  down-triangles.
}
\end{figure}

It is natural to ask why the Fig.\,\ref{F3TOTEM}\,-\,Image~3 SPM result is meaningfully different from the other three.  This issue is addressed by Fig.\,\ref{F5TOTEM}, which compares the Method~A extrapolation of TOTEM data, detailed in Sect.\,\ref{MethodA}, with the SPM interpolation of ISR and TOTEM data explained in Sect.\,\ref{MethodAISR}.  Evidently, the latter procedure produces a much noisier cross-section, with strongly overlapping $|t|$-bins on $|t|\geq 0.68\,$GeV$^2$.  These features can be traced to the character of the ISR data, whose precision does not match that of the TOTEM measurements.  Looking carefully at the ISR data \cite[Fig.\,2]{Amaldi:1979kd}, one finds that the characteristic points, which are an integral part of Method~A, are not well separated in the ISR data and their location does not show clear ordering.  In fact, neither the depths of the diffractive minima nor their locations show a uniform evolution with energy.  This is not an issue for the Method~B approach, Sect.\,\ref{MethodBISR}, because it works directly with the D0 $|t|$-bins, which are far removed from the diffractive minima of the ISR cross-sections.

\section{Caveat owing to the nature of D0 data}
%
It is worth taking a closer look at the SPM-average comparison with D0 data in Fig.\,\ref{F4TOTEM}.  Relative to all other data, the point at $|t|=0.8\,$GeV$^2$ is unexpectedly low.  If the downward displacement is real, then it suggests a second minimum in the cross-section.  Such a feature is not anticipated.  It is therefore worth considering the effect this point has on the significance of the difference between the  SPM predictions for the $\surd s = 1.96\,$TeV $pp$ elastic differential cross-sections and the D0 $p\bar p$ cross-section.  This is especially true given that any odderon signal is expected to be concentrated in the neighbourhood of the known diffractive minimum and the $|t|=0.8\,$GeV$^2$ point lies beyond that zone.

Omitting the $|t|=0.8\,$GeV$^2$ D0 point and computing the $\chi^2$, then with six degrees-of-freedom one finds $p=3.1$\%; thus, the D0 and SPM analyses of ISR and TOTEM cross-sections disagree with significance
\begin{equation}
\label{SigodderonExcluded}
{\rm S}^{\rm A+B+B_{+{\rm ISR}}}_{{\mathbb O},\,|t|\neq 0.8\,{\rm GeV}^2} = 2.2\sigma\,.
\end{equation}

\section{Summary and conclusion}
\label{epilogue}
Our multifaceted SPM analysis has enabled a comparison between the D0 $p\bar p$ elastic differential cross-section and function-form-unbiased extrapolations based on kindred TOTEM $pp$ measurements \cite{TOTEM:2018psk, TOTEM:2011vxg, TOTEM:2015oop, TOTEM:2018hki} and, uniquely, interpolations based on a combination of TOTEM and ISR $pp$ measurements \cite{Amaldi:1979kd}.  The comparisons provide evidence in support of the claim that $t$-channel exchange of a family of colourless, crossing-odd states -- the odderon -- is necessary in order to describe high-energy elastic scattering, with significance ${\rm S}_{\mathbb O} = (2.2 - 2.6)\sigma$.  In arriving at this position, we have independently confirmed the qualitative conclusions drawn using function-form-dependent interpolations and extrapolations in Ref.\,\cite{TOTEM:2020zzr}.

The next step, \emph{viz}.\ that to a claim for discovery of the odderon, relies on the combination of the evidence revealed herein with inferences derived from comparisons between different cross-section measurements \cite{TOTEM:2017sdy} and associated odderon-excluding model predictions.  Such comparisons are reported \cite{TOTEM:2020zzr, COMPETE:2002jcr} to deliver combined significances in the range $(3.4 - 4.6)\sigma$.  Accepting those reports at face value and using Eqs.\,\eqref{Sigodderon}, \eqref{SigodderonExcluded}, then we arrive at a final significance in the range \cite{Bityukov:2008zz}
\begin{equation}
\label{SigodderonF}
{\rm S}_{\mathbb O} = (4.0 - 5.2)\sigma\,.
\end{equation}
This being the case, then somewhat more evidence is required before one may definitively claim experimental observation of the odderon.

\medskip
\noindent\emph{Acknowledgments}.
We are grateful for constructive comments from V.~Mokeev and J.~Rodr\'{\i}guez-Quintero.
Use of the computer clusters at the Nanjing University Institute for Nonperturbative Physics is gratefully acknowledged.
Work supported by:
National Natural Science Foundation of China (grant no.\,12135007);
Natural Science Foundation of Jiangsu Province (grant no.\ BK20220323);
and STRONG-2020 ``The strong interaction at the frontier of knowledge: fundamental research and applications'' which received funding from the European Union's Horizon 2020 research and innovation programme (grant no.\,824093).
%



\end{document}